\documentclass[a4paper]{aipproc}
\layoutstyle{6x9}

\def\NPA{{\em Nucl. Phys.} {\bf A}}

\def\PLB{{\em Phys. Lett.} {\bf B}}

 \def\tstrut{\vrule height2.5ex depth0pt width0pt} 
  
\begin{document}

\title 
      [Kaonic Atoms]
      {Optical Potentials in Kaonic Atoms}


\author{Carmen Garc\'{\i}a-Recio}{
  address={Departamento F\'{\i}sica Moderna, University of Granada, E-18071
  Granada, Spain},
  email={g_recio@ugr.es}
}
\iftrue
\author{Juan Nieves}{
  address={Departamento F\'{\i}sica Moderna, University of Granada, E-18071
  Granada, Spain},
  email={jmnieves@ugr.es}
}
\author{Eulogio Oset}{
  address={Departamento de F\'{\i}sica Te\'orica and IFIC, Centro Mixto
Universidad de Valencia-CSIC,
Paterna, Aptdo. Correos 22085, 46071 Valencia, Spain} 
}
\author{Angels Ramos}{
  address={Departament
d'Estructura i Constituents de
la Mat\`eria, Universitat de Barcelona, 08028 Barcelona, Spain }
}
\fi

\copyrightholder{Carmen Garc\'{\i}a Recio}
\copyrightyear  {2001}

\begin{abstract}
  A microscopic optical potential based on a chiral model is used as
   a starting point for studying kaonic atoms levels. We add to this
   potential a phenomenological part fitted to the experimentally
   known shifts and widths of kaonic levels. This fitted potential is
   used to predict deeply bound atomic levels, as well as nuclear
   levels. Comparison with the predictions of other optical models
   found in the literature is done. Also the effects on the kaonic
   atoms levels of certain known non-local contributions to the
   optical potential are also analyzed.
\end{abstract}
\date{\today}
\maketitle

\section{The selfconsistent $K^-$ selfenergy of Ramos-Oset}
\label{sec:RO}
 The problem of kaonic atoms has regained interest recently. First, due 
to the new perspective that the use of
 chiral Lagrangians has brought into the problem \cite{ram}. Second,
 because of the need to obtain
 accurately the kaon selfenergy in a nuclear medium, in view of the possibility to get
 kaon condensates in neutron-proton stars.  Third, the interpretation
 of the enhancement of the $K^-$ yields in heavy ion reactions 
  relies on the value of
 the $K^-$ selfenergy in the nuclear medium.

\begin{figure}
  {\includegraphics[height=.245\textheight]{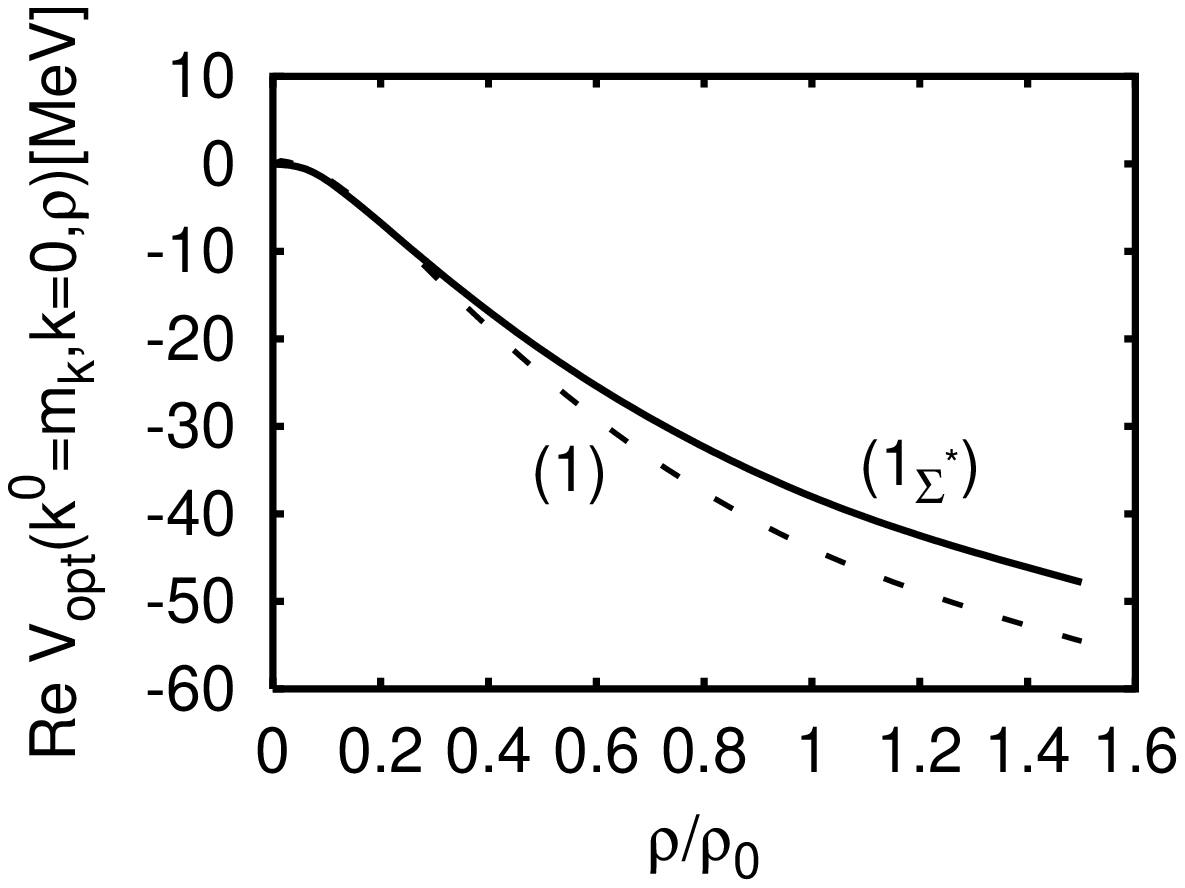}}
  {\includegraphics[height=.245\textheight]{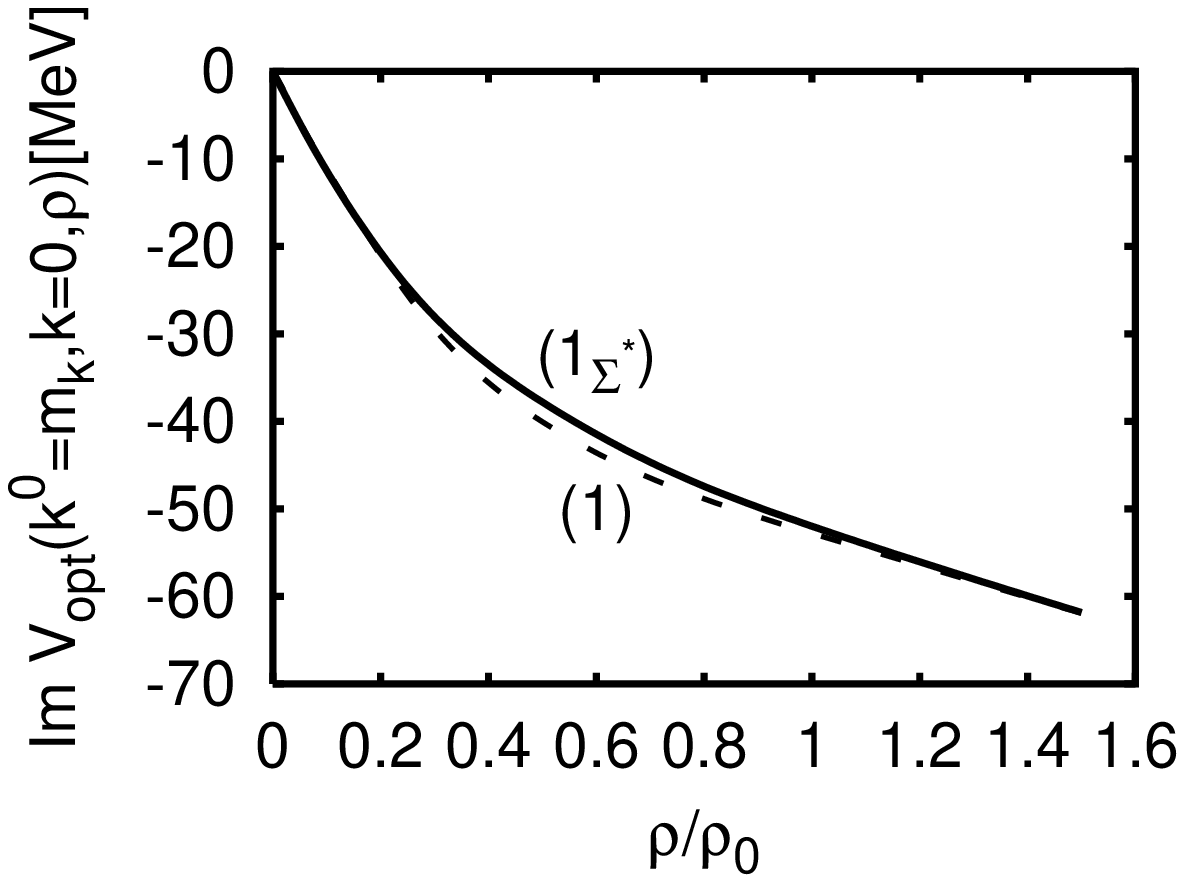}}
  \caption{$K^-$ optical potentials versus nuclear density.}
\label{fig:vopt}
\end{figure}

  The dominance of the s--wave in the
 elementary $\bar{K} N$ interaction  has been the justification for using 
 traditionally s--wave $\bar{K}$  
 nucleus optical potentials 
 \cite{Friedman99,baca}, by means
 of which good agreement with data can be obtained.  
In the work of Ramos-Oset~\cite{ram},   
the s--wave self-energy $\Pi ^s_{\bar K}$ of the $K^-$ meson in nuclear
matter is calculated in a
selfconsistent microscopic approach, using an in medium effective $\bar
K N$  interaction, $t^{eff}_{ij}$, obtained from the lowest-order
meson-baryon chiral lagrangian $V_{ij}$ by solving the coupled-channel
Bethe-Salpeter equation for the meson-baryon sector with strangeness S=$-1$.
For our calculations of shifts and widths in kaonic atoms, we take as
starting point this local theoretical potential, $V_{\rm opt}^{(1)}$ 
(dashed line in fig.~\ref{fig:vopt}). We also use an improved version
of this potential, 
$V_{\rm opt}^{(1_{\Sigma^*})}$ (shown in fig.~\ref{fig:vopt} with
solid line),  
which includes the $\Sigma^*h$--excitation in the selfconsistent calculation of 
the s--wave $K^-$ optical potential.
\section{Calculation of shifts and widths of kaonic atoms}
{\bf \ \ \ Purely s--wave potential.}
By using the s--wave potentials (1) and (1$_{\Sigma^*}$) we calculate 
the shifts and widths corresponding to 63 experimental data (see
table~\ref{tab:bind-gap1}) 
obtaining values of $\chi^2$ per data of 3.76 and 2.89, respectively.  
The agreement with the data is quite satisfactory if one takes into account that the 
potentials are purely theoretical without any free parameter. \\

{\bf p--wave.}
The lowest order p--wave optical potential~\cite{gar01} includes 
the lowest order p--wave chiral lagrangian plus the contribution of the 
$\Lambda h$, $\Sigma h$ and $\Sigma^* h$ excitations. Using the s--wave potential 
$V_{\rm opt}^{(1)}$ plus this p--wave part to solve the Klein-Gordon equation, we 
find the results of row (2) in table~\ref{tab:bind-gap1}, with 
$\chi^2/N = 4.00$. 
We observe that the change in the shifts and widths due to the p--wave are much  
smaller than the experimental uncertainties. \\

{\bf s--wave induced non-local effects.}
The s--wave optical potential  of a $K^-$ in a nuclear medium of
density $\rho$ depends on 
the $K^-$ energy $\omega$ and momentum $\vec{k}$.
For kaonic atoms, the potentials (1) and (1$_{\Sigma^*}$) were
evaluated at threshold ($\omega = m_{K}, \vec{k}=0$). Now, we
consider the corrections to the optical potential due to the explicit  
$\omega$ and $\vec{k}$ dependence. Expanding the $K^-$ selfenergy at
first order around threshold:
\begin{equation}
\label{eq:nonloc}
\Pi(\omega,\vec{k},\rho)=2\omega V_{\rm
  opt}=\Pi(m_K,0,\rho)+b(\rho){\vec{k}}^2 + c(\rho) (\omega-m_K)
\end{equation}
The momentum $\vec{k}$ is not defined for a $K^-$ bound in an atom
and instead it becomes an operator. Having evaluated the selfenergy in
nuclear matter, it is not well defined which kind of $\vec{\nabla }
\vec{\nabla }$ operator will correspond to the factor $\vec{k}^2$.
So different realizations for the operator $\vec{k}^2$ have been considered, see
rows (3a) and (3b) in table~\ref{tab:bind-gap1}. The effect of considering
the $c (\omega-m_K)$ term is shown in row (4) of the table.
We observe that the effect of any of these non local terms on shifts
and widths are smaller that the error bars of the experimental data.
For more details on non-local effects see ref.~\cite{gar01}.
{\small
\begin{table}
\vspace*{-.25cm}
\label{tab:bind-gap1}
\begin{tabular}{l||r|rr|rr|rr|rr|rr}
 \  &$\chi^2/N$
&\multicolumn{2}{ c|} {$_5^{10}$B}
&\multicolumn{2}{|c|} {$_{13}^{27}$Al}
&\multicolumn{2}{|c|} {$_{29}^{63}$Cu}
&\multicolumn{2}{|c|} {$_{48}^{112}$Cd}
&\multicolumn{2}{|c } {$_{92}^{238}$U}
\\
 &  
 & $-\epsilon_{2p}$ & $\Gamma_{2p}$
 & $-\epsilon_{3d}$ & $\Gamma_{3d}$
 & $-\epsilon_{4f}$ & $\Gamma_{4f}$
 & $-\epsilon_{5g}$ & $\Gamma_{5g}$
 & $-\epsilon_{7i}$ & $\Gamma_{7i}$
 \\\hline\hline\tstrut 
(1) & 3.76
& 217 &  551
& 109 &  368
& 384 & 1121
& 528 & 1437
& 330 & 1090
\\\hline
(2)   & 4.00
& 213 &  542
& 110 &  362
& 392 & 1110
& 543 & 1420
& 350 & 1076
\\\hline
(3a) & 3.20
& 211 &  565
& 102 &  397
& 361 & 1229
& 494 & 1588 
& 302 & 1291
\\\hline
(3b) & 4.00
& 234 &  564 
& 118 &  383
& 415 & 1172
& 568 & 1515
& 357 & 1196
\\\hline
(4) & 3.69
& 217 &  552
& 110 &  371
& 388 & 1141
& 534 & 1465
& 337 & 1166
\\\hline\hline
(1$_{\Sigma^*}$) & 2.89
& 208 &  575
& 105 &  398
& 373 & 1219
& 512 & 1550
& 320 & 1201
\\\hline\hline
(1m) & 1.52
& 159 &  742
&  69 &  438
& 335 & 1290
& 490 & 1610
& 270 & 1100
\\\hline
(1$_{\Sigma^*}b_0$) & 1.30
& 156 &  722
&  68 &  449
& 309 & 1368
& 453 & 1732
& 255 & 1241
\\\hline
(1$_{\Sigma^*}B_0$) & 1.41
& 148 &  673
&  70 &  438
& 308 & 1386
& 448 & 1781
& 279 & 1297
\\\hline\hline
Exp & - 
& 208 &  810
&  80 &  443
& 370 & 1370
& 400 & 2010
& 260 & 1500
\\  
   &
& $\pm$35 & $\pm$100
& $\pm$13 & $\pm$22
& $\pm$47 & $\pm$170
& $\pm$100 & $\pm$440
& $\pm$400 & $\pm$750
\end{tabular}
\caption{ Widths and shifts of representative kaonic 
atom levels in 
eV obtained from different potentials. Also $\chi^2$ per number of data, 
$N=63$, is shown. 
Row (1) correspond to
the local potential of ref.~\protect\cite{ram}. 
Rows (2) to (4) correspond to different non-local additions to this dominant piece (1):\protect\\
\hspace*{0.4cm}
(2) Only p--wave non-local effects due to the coupling of  $K^- N$ to  $ \Lambda$, $\Sigma$ and  $\Sigma^*$, are added.\protect\\ 
\hspace*{0.4cm}
(3) Only lowest order in momentum $b\,
\vec{q}\,^2$ non-local terms of the s--wave potential, see eq.~(\protect\ref{eq:nonloc}), are included
in two different ways: (3a) 
$-\vec\nabla b\vec\nabla$, (3b) $-\vec\nabla b\vec\nabla -0.5 (\Delta b)$.\protect\\
\hspace*{0.4cm}
(4) Only energy dependent $c\, (\omega-\mu)$ ``non-local''
effects of eq.~(\protect\ref{eq:nonloc}) are added.\protect\\
The results of row $(1_{\Sigma^*})$ are obtained from a theoretical 
$s$--wave optical potential, like in row (1), but including 
$\Sigma^* h$ excitations.
Potentials (1m), (1$_{\Sigma^*}b_0)$ and (1$_{\Sigma^*}B_0)$ are best--fit local potentials.
\vspace*{0.75cm}    }
\end{table}
}

\section{Fits to atomic data} 
\label{sec:fits}
By adding a fitted correction $\delta V^{\rm  fit}$ to the previous
theoretical optical potentials, we achieve theoretically founded
phenomenological potentials that describe the experimental data quite
satisfactorily.  
We consider two different forms for the fitted part:\\
$2\mu~\delta V^{\rm  fit}_{b_0}(r)~=~-4\pi~(1+\mu/M_N)~\rho(r)~\delta b_0$,
with one complex parameter $\delta b_0$, and \\
$2\mu~\delta V^{\rm  fit}_{B_0}(r)~=~-4\pi~(1+\mu/M_N)~\rho(r)~(\rho(r)/\rho_0)^{1/3}~(~i~
\delta {\rm Im}B_0)$, with one real parameter $\delta{\rm
  Im}B_0$. The quantity $\mu$ is the $K^-$-nucleus reduced mass. 
The phenomenological potential (1m) is obtained from (1) plus $\delta
V^{\rm fit}_{b_0}$ with $\delta b_0 = (0.078-i0.25)$~fm. The potential
(1$_{\Sigma^*}b_0$), from (1$_{\Sigma^*}$) plus  $\delta
V^{\rm fit}_{b_0}$ with $\delta b_0 = (0.0750-i0.200)$~fm. And  the potential
(1$_{\Sigma^*}B_0$), from (1$_{\Sigma^*}$) plus $\delta
V^{\rm fit}_{B_0}$ with $\delta {\rm Im}B_0 = -0.260$~fm.
They provide good fits with $\chi^2$ per data of about 1.4, 
see rows (1m), (1$_{\Sigma^*}b_0$) and (1$_{\Sigma^*}B_0$) of
table~\ref{tab:bind-gap1}.\\

\begin{figure}
{\includegraphics[height=7.9cm]{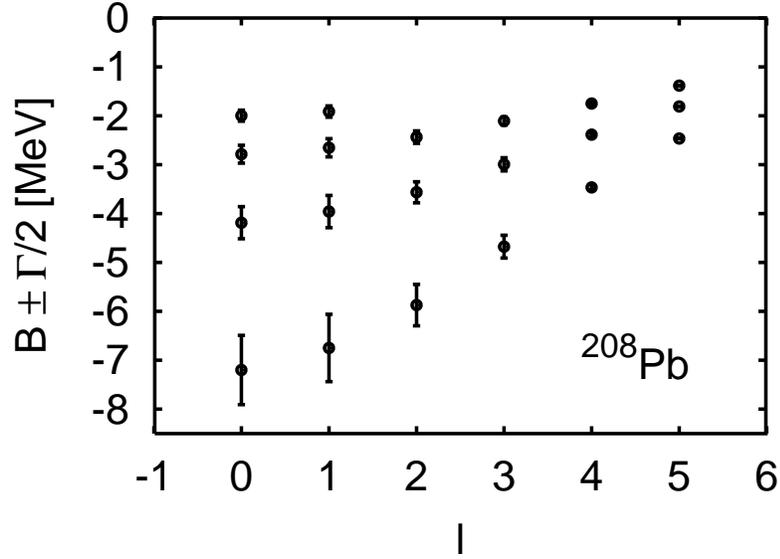}}
\caption { Binding energies $B $ of deeply bound atomic levels in
$^{208}$Pb versus angular momentum $l$. The error bar stands for
the full width $\Gamma$ of each level. They have
been computed using the $V^{\rm(1m)}_{\rm opt}$  potential.}
\label{fig:atom}
\end{figure}

{\bf Predictions: deeply--bound atomic levels.}
The fitted potential (1m), described above, is used to predict binding
energies and widths of deeply bound atomic states, not yet
observed. They are shown, for $^{208}$Pb, in fig.~\ref{fig:atom}. See
ref.~\cite{baca} for other nuclei.
One can see that the levels, including the widths, do not overlap for
a given angular momentum. 
%
%
%
%
\section{Conclusions}
\begin{itemize}
\item The selfconsistent microscopic approach  based on the chiral lagrangian 
(1$_{\Sigma^*}$) 
is quite good: $\chi^2/N = 2.9$, for 63 atomic data, with no free parameter 
in the model.
\item Non-local effects, associated to proper p--wave contributions or to momentum 
and energy dependence of the s--wave self-energy, are negligible at this stage, 
because their effect on shifts and widths 
are smaller than the current uncertainties of data.
\item An improved fitted potential, (1$_{\Sigma^*}b_0$), provides a very good fit with 
$\chi^2/63 = 1.3$ .
\item Deeply-bound  kaonic atom levels are narrow and separable, so subjected to experimental 
observation via nuclear reactions. The widths of these levels are
sensitive to different potentials by about a 20\%.
\end{itemize}
\begin{theacknowledgments}
This work was partially supported by DGICYT contract
PB98-1367 and Junta de Andaluc\'{\i}a under grant FQM 0225
\end{theacknowledgments}
%


\end{document}